\documentclass[aps,pre,twocolumn,superscriptaddress]{revtex4-1}
\usepackage{graphicx}
\usepackage{array}

\begin{document}

\title{Co-evolution of the mitotic and meiotic modes of eukaryotic cellular
division}

\author{Valmir C. Barbosa}
\email[]{valmir@cos.ufrj.br}
\affiliation{Programa de Engenharia de Sistemas e Computa\c c\~ao, COPPE,
Universidade Federal do Rio de Janeiro,
Caixa Postal 68511, 21941-972 Rio de Janeiro - RJ, Brazil}

\author{Raul Donangelo}
\affiliation{Instituto de F\'\i sica,
Universidade Federal do Rio de Janeiro,
Caixa Postal 68528, 21941-972 Rio de Janeiro - RJ, Brazil}
\affiliation{Instituto de F\'\i sica, Facultad de Ingenier\'\i a,
Universidad de la Rep\'ublica,
Julio Herrera y Reissig 565, 11.300 Montevideo, Uruguay}

\author{Sergio R. Souza}
\affiliation{Instituto de F\'\i sica,
Universidade Federal do Rio de Janeiro,
Caixa Postal 68528, 21941-972 Rio de Janeiro - RJ, Brazil}
\affiliation{Instituto de F\'\i sica,
Universidade Federal da Bahia,
Campus Universit\'ario de Ondina, 40210-340 Salvador - BA, Brazil}
\affiliation{Departamento de F\'\i sica, ICEx,
Universidade Federal de Minas Gerais,
Av.\ Ant\^onio Carlos, 6627, 31270-901 Belo Horizonte - MG, Brazil}

\begin{abstract}
The genetic material of a eukaryotic cell (one whose nucleus and other
organelles, including mitochondria, are enclosed within membranes) comprises
both nuclear DNA (ncDNA) and mitochondrial DNA (mtDNA). These differ markedly in
several aspects but nevertheless must encode proteins that are compatible with
one another for the proper functioning of the organism. Here we introduce a
network model of the hypothetical co-evolution of the two most common modes of
cellular division for reproduction: by mitosis (supporting asexual reproduction)
and by meiosis (supporting sexual reproduction). Our model is based on a random
hypergraph, with two nodes for each possible genotype, each encompassing both
ncDNA and mtDNA. One of the nodes is necessarily generated by mitosis occurring
at a parent genotype, the other by meiosis occurring at two parent genotypes. A
genotype's fitness depends on the compatibility of its ncDNA and mtDNA. The
model has two probability parameters, $p$ and $r$, the former accounting for the
diversification of ncDNA during meiosis, the latter for the diversification of
mtDNA accompanying both meiosis and mitosis. Another parameter, $\lambda$, is
used to regulate the relative rate at which mitosis- and meiosis-generated
genotypes are produced. We have found that, even though $p$ and $r$ do affect
the existence of evolutionary pathways in the network, the crucial parameter
regulating the coexistence of the two modes of cellular division is $\lambda$.
Depending on genotype size, $\lambda$ can be valued so that either mode of
cellular division prevails. Our study is closely related to a recent hypothesis
that views the appearance of cellular division by meiosis, as opposed to
division by mitosis, as an evolutionary strategy for boosting ncDNA
diversification to keep up with that of mtDNA. Our results indicate that this
may well have been the case, thus lending support to the first hypothesis in the
field to take into account the role of such ubiquitous and essential organelles
as mitochondria.

\end{abstract}

\maketitle

\section{Introduction}
\label{sec:intro}

A cell is said to be eukaryotic if it has a nucleus as well as other organelles
enclosed in membranes providing separation from the cellular medium. With one
single exception known to date \cite{kvetal16}, these other organelles include
mitochondria, the cell's powerhouses. Every multicellular organism is a
eukaryote, and so are numerous unicellular organisms as well, such as
unicellular algae and fungi. A eukaryote's genotype comprises the genetic
material found in both its nuclear DNA (ncDNA) and its mitochondrial DNA
(mtDNA). Both types of DNA are essential for the proper functioning of the cell,
so despite the fundamental differences between ncDNA and mtDNA (such as shape,
size, multiplicity, and inheritance patterns), the proteins their genes
synthesize must be compatible with one another. In fact, it is thought that such
compatibility is key to an organism's fitness in evolutionary terms
\cite{oa12,bb12,stwt13}, as well as to regulating metabolic functions and
supporting healthy aging \cite{lpetal16}. Here we consider eukaryotic cells
exclusively.

A cell's ncDNA is organized as pairs of chromosomes. Its mtDNA, in turn, is part
of a single chromosome in each mitochondrion. This form of organization is
deeply entwined with how the organism reproduces, since it supports cellular
division (the central reproductive event at the cellular level) by either of the
two most common modalities, mitosis and meiosis. The mitotic mechanism of
division is used by an organism both for its somatic cells to multiply and for
asexual (or clonal) reproduction if such is the case. During mitosis, the cell
gets divided into two identical cells, each inheriting from the original cell
an exact copy of its ncDNA and possibly mutated copies of its mtDNA. This is not
to say that ncDNA never incurs mutations, which in fact constitute the
prevailing cause of abnormalities such as cancer \cite{tlv17}, only that such
mutations are so rare as to be negligible in normal mitosis.

The other common mechanism of cellular division, that of meiosis, is central to
the sexual reproduction of organisms. When a cell undergoes meiosis, its ncDNA
is first ``shuffled'' through recombination and mutation of the genetic material
in the paired chromosomes. Each chromosome in the resulting pair gets inherited
by one of the cells produced by the division, called a gamete. Each gamete
inherits a mutated version of the parent cell's mtDNA, just as in the case of
mitosis. The encounter of two gametes, one from each parent during sexual
reproduction, gives rise to a somatic cell of the resulting offspring, now with
the paired chromosomes restored (one chromosome from each parent). This cell's
mtDNA is in general inherited from only one of the parents (the mother). Cell
division by meiosis is a much more complex process than division by mitosis, and
as such requires substantially more time to complete.

Curiously, some organisms reproduce both asexually and sexually, depending on
environmental and other factors \cite{bvhg12,yk16}, which hints at the
possibility of a deep evolutionary past in which the modes of cellular division
were much less well-defined and coexisted much more freely. If such a past
really existed, then the events that took place in it must have lied at the very
roots of the evolution of sex and of meiosis as the currently prevalent mode of
cellular division for reproduction. However, in spite of the evidence we find
today in the form of organisms adopting asexual as well as sexual reproductive
strategies, a widely accepted theory of how sex evolved is still lacking, even
though proposals ranging from the purely biological \cite{ol02,hk12} to the
algorithmic and game-theoretic \cite{lp16} have been put forward.

In this paper we aim to explore, via mathematical modeling and computer
simulations, what seems to be the most recent proposal as to why sex evolved in
the first place and moreover has endured ever since \cite{hhd15,p16}. This
proposal is based on two core assumptions (cf.\ \cite{hhd15} and references
therein). Assumption~1 is that the mutation rates in mtDNA transmission during
cellular division, known to be much higher than that in ncDNA during
recombination, have been consistently high since ancient times. Assumption~2 is
that the inheritance of mtDNA from only one of the two gametes produced by
meiosis, which is the rule for all eukaryotes with very few exceptions, has all
along been constrained by natural selection and as such has also been the rule
since the beginning. What the new theory posits is that sex evolved in order for
the mutations that ncDNA undergoes during recombination to compensate for those
of mtDNA and thereby help maintain compatibility between the two. This is backed
by Assumption~1. As for Assumption~2, it is needed to prevent mtDNA from
acquiring even more variability through the recombination that could take place
if mtDNA material were inherited from both gametes. The fundamental nature of
both recombination and mutation has been expressed mathematically in a number of
occasions (cf., e.g., \cite{ckl05,pd07,s07,sh13,s18} and references therein),
but the new proposal calls for their roles to be explored during the
co-evolution of two very distinct types of DNA. Such exploration lies at the
core of the present study.

Our model is essentially a network of genotypes with accompanying dynamical
equations, each genotype occupying a node of the network and being characterized
by its abundance (how many of it there are) and by how it relates to other
genotypes. Each genotype is intended to account for both ncDNA and mtDNA, being
therefore represented by three sequences, two standing for an ncDNA pair and one
for mtDNA. Our network allows for the coexistence of two kinds of genotype. One
of them comprises genotypes produced exclusively by mitosis, via the cloning of
any other genotypes, regardless of what kind those genotypes belong to.
Genotypes of the other kind necessarily result from the process of meiosis on
the parents' side, again with no restrictions on the kind of the parent
genotypes. Clearly, then, our model allows for the free coexistence and
intermixing of both modes of cellular division.

The model also includes provisions to incorporate the Darwinian principles of
random mutations and natural selection, in the form of two probability
parameters, $p$ and $r$, and of a measure for a genotype's fitness quantifying
the compatibility of its ncDNA and mtDNA. The parameters $p$ and $r$ are meant
to reflect the inherent randomness of ncDNA recombination and mtDNA mutation.
Their role in the model is to regulate the network's density by affecting the
interconnectedness of the genotypes, and also the pace of the model's dynamics.
A third and final parameter, $\lambda$, helps account for the different
durations of mitotic and meiotic cellular division.

Our model's dynamical equations are reminiscent of the well-known quasispecies
equations \cite{e71,es77,be06}, originally introduced to model the evolution of
prebiotic molecules and RNA viruses \cite{d09,la10,mlcgm10}, and of our own
modifications thereof to incorporate network structure \cite{bds12,bds15,bds16}.
They are presented in Sec.~\ref{sec:model}, along with all other details of
the model. We then proceed with the presentation of results in
Sec.~\ref{sec:results}, discussion in Sec.~\ref{sec:disc}, and conclusions
in Sec.~\ref{sec:concl}.

\section{Model}
\label{sec:model}

\begin{figure}[t]
\includegraphics[scale=0.85]{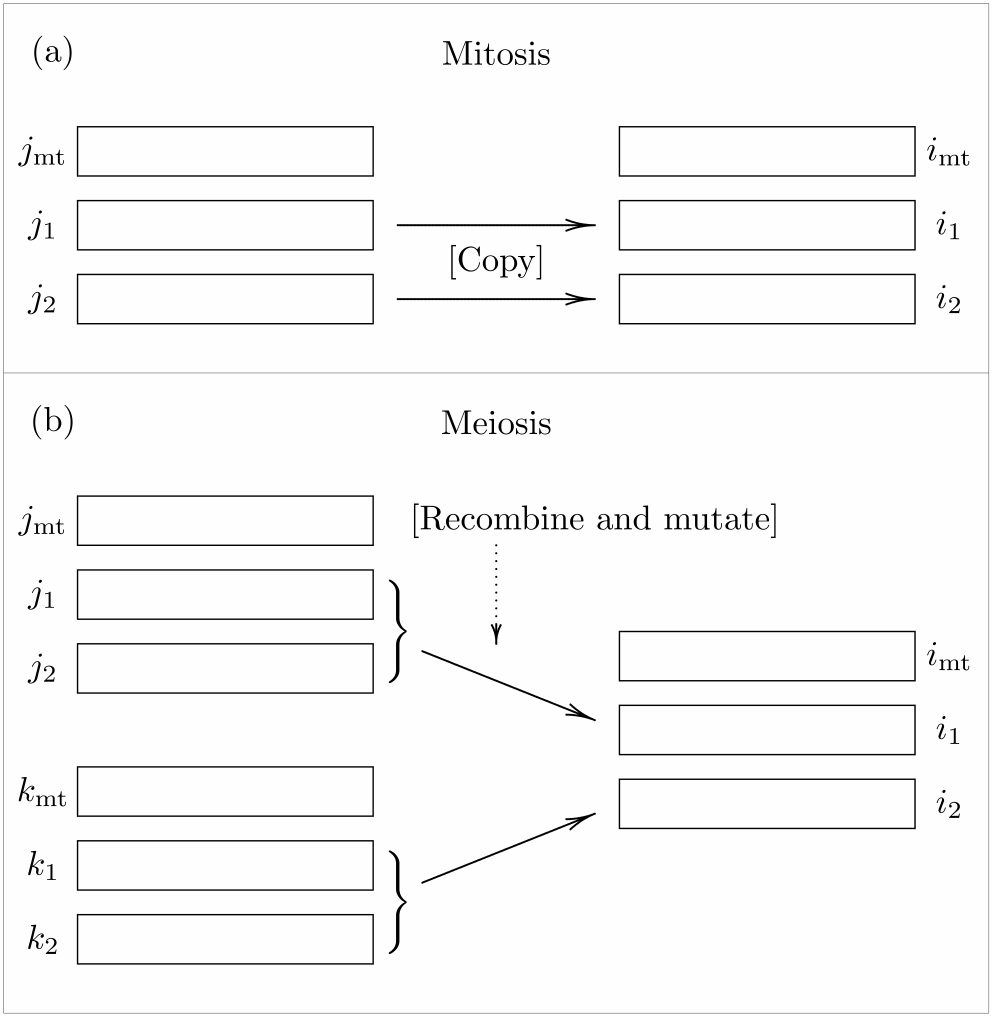}
\caption{(a) Generation of genotype $i$ by mitosis from genotype $j$. Sequences
$j_1$ and $j_2$ are copied to $i_1$ and $i_2$, respectively, while
$i_\mathrm{mt}$ is a possibly mutated copy of $j_\mathrm{mt}$. (b) Generation of
genotype $i$ by meiosis at parents $j$ and $k$. Sequence $i_1$ originates from
recombination and mutation between $j_1$ and $j_2$, sequence $i_2$ from $k_1$
and $k_2$. Sequence $i_\mathrm{mt}$ is a possibly mutated copy of either
$j_\mathrm{mt}$ or $k_\mathrm{mt}$.}
\label{fig1}
\end{figure}

We represent each genotype as a triplet of sequences, each one comprising $L$
binary digits ($0$'s or $1$'s). For genotype $i$, we denote these sequences by
$i_1$, $i_2$, and $i_\mathrm{mt}$, where $i_1$ and $i_2$ are the two chromosomes
of ncDNA and $i_\mathrm{mt}$ is the single chromosome of mtDNA. This particular
choice for the representation of a genotype incurs two strong simplifications,
viz., that each allele (gene variant) gets reduced to only two possibilities and
that the length of the mitochondrial chromosome is the same as that of the two
nuclear chromosomes (whereas, in fact, it should be orders of magnitude
shorter). As will become apparent, these simplifications have to do with
practical limitations regarding the number of distinct genotypes given $L$,
henceforth denoted by $G$, as well as the size of a network capable of
accommodating twice as many genotypes. The value of $G$ can be calculated by
first noting that, for each of the $2^L$ possibilities for $i_\mathrm{mt}$,
there are ${2^L\choose 2}$ possibilities for the (unordered) pair $i_1,i_2$ if
$i_1\neq i_2$ and $2^L$ possibilities if $i_1=i_2$. That is,
\begin{equation}
G=2^L\left[{2^L\choose 2}+2^L\right]=2^{3L-1}+2^{2L-1}.
\label{eq:G}
\end{equation}
A schematic representation of mitosis and meiosis as they act on such sequences
is given in Fig.~\ref{fig1}.

\subsection{Fitness of a genotype}

Such simplifications also facilitate the definition of a genotype's fitness in
terms of how compatible its ncDNA and mtDNA are. For genotype $i$, its fitness,
denoted by $f_i$, is given by $f_i=2^{-d_i}$, where $d_i$ is the number of loci
at which $i_\mathrm{mt}$ differs from both $i_1$ and $i_2$. Clearly, fitnesses
range from $2^{-L}$ (when $i_1$ and $i_2$ are identical and $i_\mathrm{mt}$
differs from them at all loci) to $1$ (when $i_1$ and $i_2$ do not necessarily
equal each other at all loci but $i_\mathrm{mt}$ coincides with them wherever
equality happens).

Later in the sequel it will become handy for us to know how many distinct
genotypes there exist for which fitness equals some fixed value $2^{-k}$. This
number, $n_k$, can be easily calculated, as follows. Let $\alpha$ be the number
of loci at which $i_1$ and $i_2$ differ and $\beta\ge k$ be the number of loci
at which $i_1$ and $i_2$ are equal. Clearly, $\alpha+\beta=L$. If the partition
between the $\alpha$ and the $\beta$ loci were fixed, then the value of $n_k$,
which depends on whether $\alpha=0$ or $\alpha>0$, would be given as follows.
For $\alpha=0$, $n_k$ would equal simply ${L\choose k}2^L$. For $\alpha>0$, it
would equal ${\beta\choose k}2^{2\alpha+\beta-1}$, where $2^\alpha$
possibilities for $i_1$ (hence for $i_2$) and $2^\alpha$ for $i_\mathrm{mt}$ are
accounted for on the $i_1\neq i_2$ side of the partition, as well as $2^\beta$
possibilities for $i_1$ (hence for $i_2$) on the $i_1=i_2$ side, and moreover
such total number of possibilities gets divided by $2$ to account for the fact
that swapping $i_1$ and $i_2$ does not change genotype $i$. Letting the
partition vary yields
\begin{eqnarray}
n_k
&=&{L\choose k}2^L+
\sum_{\alpha=1}^{L-k}{L\choose\alpha}{L-\alpha\choose k}2^{\alpha+L-1}\cr
&=&2^{L-1}{L\choose k}
\left[1+\sum_{\alpha=0}^{L-k}{L-k\choose\alpha}2^\alpha\right]\cr
&=&2^{L-1}{L\choose k}(1+3^{L-k}).
\label{eq:nk}
\end{eqnarray}
As expected, $\sum_{k=0}^Ln_k=G$. Moreover, the sum total of all genotypes'
fitnesses, henceforth denoted by $F$, is given by
\begin{equation}
F=\sum_{k=0}^L n_k2^{-k}=\frac{3^L+7^L}{2}.
\label{eq:F}
\end{equation}

\subsection{A note on hypergraphs}

Most network models rely on graphs as the natural means of representation. A
graph is defined simply as a set $N$ of nodes and a set of edges that can be any
subset of $N\times N$, so clearly an edge is defined by the pair of nodes it
interconnects. Such a pair is unordered for undirected graphs, ordered for
directed graphs. Given the ordered pair $(i,j)$, the edge it stands for is said
to be directed from node $i$ to node $j$.

As it turns out, however, this representational scheme is not entirely adequate
in the present case, owing mainly to the need to represent not only the
generation of genotypes by mitosis but also the generation that results from
meiosis on the parents' side. To this end, we resort to the generalization of
graphs known as hypergraphs \cite{b89}. A hypergraph shares with a graph the
fact of being defined on a set $N$ of nodes, but differs from a graph in that
the means it employs to represent an interconnection, now called a hyperedge, is
more general than an edge. At the level of generality that we require in this
paper, a hyperedge is a nonempty multiset with elements from $N$.

Just as in the case of graphs, directed hypergraphs can also be considered
\cite{glpn93}. In a directed hypergraph, a hyperedge is partitioned into node
multisets $S$ and $D$, called the hyperedge's source and destination node
multisets, respectively. Our use of hypergraphs will include directed hyperedges
like $\{i,j\}$ with $S=\{j\}$ and $D=\{i\}$ (really the same as an edge directed
from $j$ to $i$ in a graph), possibly with $i=j$, and directed hyperedges like
$\{i,j,k\}$ with $S=\{j,k\}$ and $D=\{i\}$, possibly with any of the node
repetitions $i=j$, $i=k$, $j=k$, or $i=j=k$.

For $i\in N$, we denote the set of all source multisets $S$ for which the
hyperedge directed from $S$ to $D=\{i\}$ exists by $I_i$ (the input set to node
$i$). We write either $j\in I_i$ or $jk\in I_i$ for the two possible types of
hyperedge in our case. Correspondingly, we denote the output set of node $j$ by
$O_j$ and that of the node pair $j,k$ by $O_{jk}$, writing $i\in O_j$ and
$i\in O_{jk}$, respectively. In these notations, the use of $jk$ refers to an
unordered pair of genotypes.

\subsection{Network structure}

\begin{figure}[t]
\includegraphics[scale=0.85]{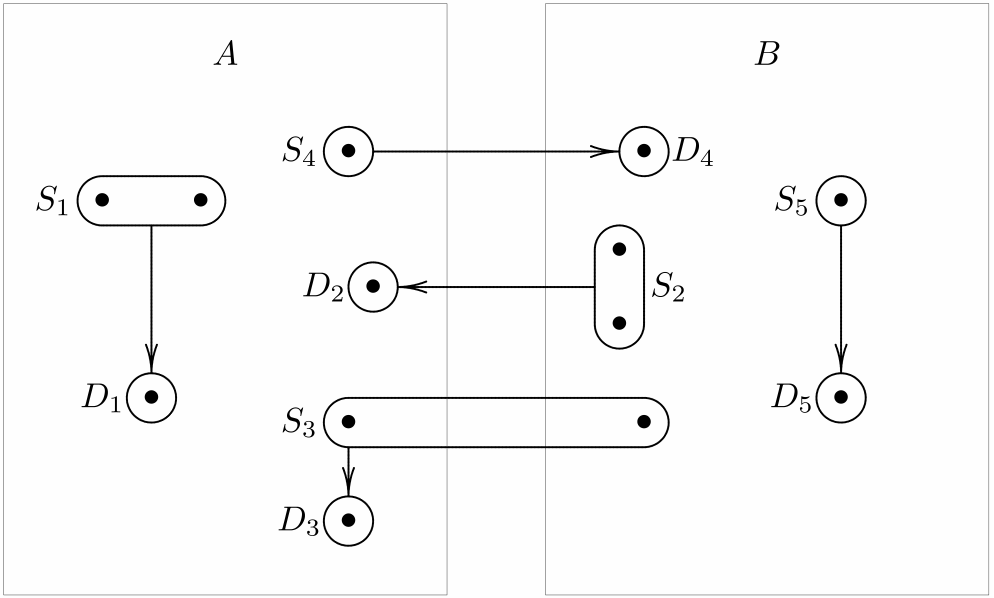}
\caption{Set $A$ contains meiosis-generated genotypes, $B$ mitosis-generated
genotypes. Three hyperedges, $S_1\to D_1$, $S_2\to D_2$, and $S_3\to D_3$,
indicate genotype generation by meiosis, two from same-set parents, one from
mixed parents. Two other hyperedges, $S_4\to D_4$ and $S_5\to D_5$, indicate
genotype generation by mitosis, the former from a genotype in $A$, the latter
from another genotype in $B$.}
\label{fig2}
\end{figure}

Our network of interacting genotypes is represented by a directed hypergraph $H$
whose set of nodes $N$ has two nodes for each of the $G$ possible genotypes. We
partition $N$ into sets $A$ and $B$, each with a full set of distinct genotypes.
That is, any possible genotype is represented in $H$ by a node in $A$ and
another in $B$. What distinguishes these two nodes is the set of hyperedges
directed toward them: nodes in $B$ are meant to represent genotypes generated by
mitosis, so $I_i$ for $i\in B$ contains single-node sets only; nodes in $A$ are
for genotypes generated by meiosis of their parents, so $I_i$ for $i\in A$
contains two-node multisets only. In either case, the nodes that go into the
sets of $I_i$ originate from the entirety of $N$, regardless of the partition
into $A$ and $B$. That is to say, it is possible for a genotype in $B$ to
originate by mitosis from either a genotype in $A$ or one in $B$. In the same
vein, each of the two genotypes that undergo meiosis to give rise to a genotype
in $A$ can be a member of $A$ or $B$. This is illustrated in Fig.~\ref{fig2}.

The description of hypergraph $H$ is completed by specifying its hyperedges. We
do this based on two probability parameters, $p$ and $r$, which regulate the
occurrence of mutation that accompanies ncDNA recombination and that which mtDNA
undergoes, respectively. Probability $p$, therefore, affects the expected number
of hyperedges incoming to nodes in $A$, whereas probability $r$ has a similar
effect on nodes in both $A$ and $B$. Clearly, then, the resulting $H$ is to be
regarded as a sample of the random hypergraph defined by the probabilities $p$
and $r$ on set $N$. By proceeding in this way, we allow for much greater
variation regarding genotype interaction.

Henceforth, we let $h(s,t)$ denote the Hamming distance between sequences $s$
and $t$ and $R(s,t)$ denote the set of all sequences that can result from
recombination between $s$ and $t$. Each sequence in $R(s,t)$, therefore, equals
both $s$ and $t$ at $L-h(s,t)$ loci and equals either one or the other at the
remaining $h(s,t)$ loci. Therefore, $R(s,t)$ contains $2^{h(s,t)}$ sequences.

\subsection{Mitotic hyperedges}

A mitotic hyperedge is directed from $S=\{j\}$ to $D=\{i\}$ with $j\in A\cup B$
and $i\in B$, possibly with $i=j$. This hyperedge can only exist if genotypes
$i$ and $j$ have the same ncDNA (i.e., both $i_1=j_1$ and $i_2=j_2$), since
normal division by mitosis does not alter the nuclear genetic material. If this
holds, then the hyperedge exists with probability $p_{j,i}$, given by
\begin{equation}
p_{j,i}=r^{h(i_\mathrm{mt},j_\mathrm{mt})}.
\label{eq:pmit}
\end{equation}
That is, the existence of the hyperedge depends on how likely it is for the
mtDNA of genotype $j$ to mutate into that of genotype $i$. If the two are
identical (hence $i=j$), then $p_{j,i}=1$. Consequently, every genotype in $B$
has at least two incoming hyperedges, one originating from itself and another
originating from its identical counterpart in $A$.

\subsection{Meiotic hyperedges}

A meiotic hyperedge has $S=\{j,k\}$ and $D=\{i\}$ with $j,k\in A\cup B$ and
$i\in A$, allowing for $i=j$, $i=k$, $j=k$, or $i=j=k$. Unlike its mitotic
counterpart, a meiotic hyperedge is in no way constrained by how the genetic
material of the genotypes involved relate to one another. Its existence is
therefore unconditionally random and occurs with probability $p_{jk,i}$, whose
calculation must take into account the recombination of genetic material of
both $j$ and $k$, possibly affected by mutation, and the inheritance by genotype
$i$ of a mutated version of the mtDNA of either $j$ or $k$. Given all the
independences involved, $p_{jk,i}$ can be expressed as
\begin{equation}
p_{jk,i}=\pi_{jk,i}\rho_{jk,i},
\label{eq:pmei}
\end{equation}
where $\pi_{jk,i}$ is the probability of the ncDNA-related events and
$\rho_{jk,i}$ is the probability of the mtDNA-related ones.

In order to calculate probability $\pi_{jk,i}$, we must take into account the
fact that, even though sequences $i_1$ and $i_2$ are interchangeable (swapping
them does not affect genotype $i$), there are two fundamentally distinct ways
they can originate from genotypes $j$ and $k$, depending on whether $i_1$ is
paired with $j$ and $i_2$ with $k$, or conversely. Thus,
$\pi_{jk,i}$ is given by
\begin{equation}
\pi_{jk,i}=
\pi_{j\to i_1}\pi_{k\to i_2}+\pi_{j\to i_2}\pi_{k\to i_1}-\pi_\mathrm{mixed},
\label{eq:pmei0}
\end{equation}
where $\pi_{j\to i_1}$ is the probability that $i_1$ results from recombination
at $j$ with the intervening effect of mutation, and similarly for
$\pi_{j\to i_2}$, $\pi_{k\to i_1}$, and $\pi_{k\to i_2}$. As for
$\pi_\mathrm{mixed}$, it is the probability that, in a population of
genotype-$i$ individuals, some have inherited $i_1$ from genotype $j$ and $i_2$
from genotype $k$ while others have inherited $i_1$ from genotype $k$ and $i_2$
from genotype $j$. This probability is counted twice as the first two terms in
Eq.~(\ref{eq:pmei0}) are added up, so subtracting it off the total is meant
to correct for this.

The probabilities appearing in Eq.~(\ref{eq:pmei0}) are made explicit by
resorting to the set $R(s,t)$ of all sequences that can result from
recombination between sequences $s$ and $t$. We do this by assuming that all
recombinations between ncDNA sequences occur without any bias toward any two of
the sequences (so every sequence in $R(s,t)$ is equiprobable) and that, in a way
similar to that of the mitotic case, the recombination-related mutation that
accompanies the transformation of sequence $s$ into sequence $t$ occurs with
probability $p^{h(s,t)}$. We then have
\begin{equation}
\pi_{j\to i_1}=2^{-h(j_1,j_2)}\sum_{\ell\in R(j_1,j_2)}p^{h(\ell,i_1)},
\label{eq:pmei1}
\end{equation}
\begin{equation}
\pi_{j\to i_2}=2^{-h(j_1,j_2)}\sum_{\ell\in R(j_1,j_2)}p^{h(\ell,i_2)},
\label{eq:pmei2}
\end{equation}
\begin{equation}
\pi_{k\to i_1}=2^{-h(k_1,k_2)}\sum_{\ell\in R(k_1,k_2)}p^{h(\ell,i_1)},
\label{eq:pmei3}
\end{equation}
\begin{equation}
\pi_{k\to i_2}=2^{-h(k_1,k_2)}\sum_{\ell\in R(k_1,k_2)}p^{h(\ell,i_2)},
\label{eq:pmei4}
\end{equation}
and
\begin{eqnarray}
\lefteqn{\pi_\mathrm{mixed}=}\hspace{0.1in}\\
&2^{-1}[&
\pi_{j\to i_1}\pi_{k\to i_2}
(\pi_{j\to i_2}+\pi_{k\to i_1}-\pi_{j\to i_2}\pi_{k\to i_1})+\nonumber \\
&&
\pi_{j\to i_2}\pi_{k\to i_1}
(\pi_{j\to i_1}+\pi_{k\to i_2}-\pi_{j\to i_1}\pi_{k\to i_2})].
\label{eq:pmei5}
\end{eqnarray}
Equation~(\ref{eq:pmei5}), in particular, can be understood by analyzing the
hypothetical case of a completely uniform population of genotype-$i$
individuals, that is, one in which every $i_1$ comes from $j$ and every $i_2$
from $k$, or conversely. The probability that this happens is
$\pi_{j\to i_1}(1-\pi_{k\to i_1})\pi_{k\to i_2}(1-\pi_{j\to i_2})+
\pi_{j\to i_2}(1-\pi_{k\to i_2})\pi_{k\to i_1}(1-\pi_{j\to i_1})$, which can be
rewritten as
$\pi_{j\to i_1}\pi_{k\to i_2}+\pi_{j\to i_2}\pi_{k\to i_1}-2\pi_\mathrm{mixed}$.
This expression is entirely consistent with that on the right-hand side of
Eq.~(\ref{eq:pmei0}), since the amount to be subtracted off
$\pi_{j\to i_1}\pi_{k\to i_2}+\pi_{j\to i_2}\pi_{k\to i_1}$ to ensure that the
probability of a mixed population is counted only once is exactly half the
amount to be subtracted to ensure uniformity.

As for probability $\rho_{jk,i}$, its expression is simply
\begin{equation}
\rho_{jk,i}=
r^{h(i_\mathrm{mt},j_\mathrm{mt})}+
r^{h(i_\mathrm{mt},k_\mathrm{mt})}-
r^{h(i_\mathrm{mt},j_\mathrm{mt})+h(i_\mathrm{mt},k_\mathrm{mt})},
\label{eq:pmei6}
\end{equation}
where the term subtracted at the end is the probability that, in a population
whose individuals all share genotype $i$, some have inherited their mtDNA from
genotype $j$, some from genotype $k$. As in the case of
Eq.~(\ref{eq:pmei0}), the probability that this happens is counted twice as
the first two terms are added up. The subtraction fixes this.

\subsection{Interpretation of the base probabilities}

As we consider Eqs.~(\ref{eq:pmit}), (\ref{eq:pmei1})--(\ref{eq:pmei4}),
and~(\ref{eq:pmei6}), it is important to note that the base probabilities used
($p$, $r$, and $2^{-1}$) can be interpreted as point probabilities affecting
each one of the loci in question equally and independently but not the others
(whose probability of being affected is $0$). Thus, if $b$ is one of the three
base probabilities and $h$ is the number of loci at which two sequences differ,
then the probability that all $h$ loci get affected is indeed the $b^h$ used in
various forms in those equations. This formulation is reminiscent of the
uniform-susceptibility model of quasispecies mutation, introduced elsewhere in
an attempt to circumvent the implausibility of certain common assumptions
(cf.\ \cite{bds12} and references therein).

\subsection{Network dynamics}
\label{netdyn}

Given hypergraph $H$, an instance of the random-hypergraph model described so
far, the resulting network dynamics is expressed by a set of $2G$ coupled
differential equations, one for each of the genotypes (nodes) in the network.
These equations give the rates at which the genotypes' abundances vary with
time and depend on the same probabilities $p_{j,i}$ (equation.~(\ref{eq:pmit}))
and $p_{jk,i}$ (Eq.~(\ref{eq:pmei})) used to sample hypergraph $H$ in the
first place. Readily, through these probabilities the parameters $p$ and $r$
affect network dynamics as much as network structure. Mutation rates in mtDNA
are usually much higher than those in ncDNA (cf., e.g., \cite{hchlck08} and
references therein), suggesting that we use $p\ll r$.

For use in the dynamics, such probabilities must be normalized so that summing
them up for all $i\in O_j$ with $j$ fixed yields $1$, and so does summing them
up for all $i\in O_{jk}$ with $j,k$ fixed. The probabilities' normalized
versions are, respectively,
\begin{equation}
q_{j,i}=\frac{p_{j,i}}{\sum_{\ell\in O_j}p_{j,\ell}}
\label{eq:qmit}
\end{equation}
and
\begin{equation}
q_{jk,i}=\frac{p_{jk,i}}{\sum_{\ell\in O_{jk}}p_{jk,\ell}}.
\label{eq:qmei}
\end{equation}

We first give the differential equations for the absolute abundances of the
genotypes. For genotype $i$, this absolute abundance is denoted by $X_i$. If
$i\in A$ (that is, genotype $i$ is the result of meiosis for parents
$jk\in I_i$), we have
\begin{equation}
\dot{X}_i=\sum_{jk\in I_i\atop j\neq k}f_jf_kq_{jk,i}\min\{X_j,X_k\}+
\sum_{jk\in I_i\atop j=k}f_j^2q_{jj,i}\frac{X_j}{2}.
\label{eq:XA}
\end{equation}
In this equation, the influence exerted on $\dot{X}_i$ by each $jk\in I_i$
depends both on the fitnesses $f_j$ and $f_k$ of the two parents and on
probability $q_{jk,i}$. It also depends on how abundant the pairs they form can
be, which in turn depends on whether $j\neq k$ or $j=k$. In the former case, the
number of possible pairs is given by $\min\{X_j,X_k\}$. In the latter, the
number is $X_j/2$.

If $i\in B$ (that is, genotype $i$ results from mitosis for $j\in I_i$), then a
simpler equation ensues,
\begin{equation}
\dot{X}_i=\lambda\sum_{j\in I_i}f_jq_{j,i}X_j,
\label{eq:XB}
\end{equation}
where the role played by each $j\in I_i$ in altering $\dot{X}_i$ is again
dependent on the fitness $f_j$, the probability $q_{j,i}$, and the abundance
$X_j$. The parameter $\lambda$ allows us to tune the entire dynamics so that
the speed with which mitosis and meiosis occur relative to each other can be
experimented with. Normally mitosis is a substantially faster process than
meiosis (cf., e.g., \cite{cwkdk00} and references therein), which to some extent
is already accounted for in Eqs.~(\ref{eq:XA}) and~(\ref{eq:XB}), owing to
the presence of squared fitnesses in the former equation. Tuning through the
value of $\lambda$ is then expected to work in conjunction with this.

Equations~(\ref{eq:XA}) and~(\ref{eq:XB}) imply the unbounded growth of every
genotype's absolute abundance. A better approach is then to turn to the
genotypes' relative abundances instead, since these are constrained to lie in
the interval $[0,1]$. The relative abundance of genotype $i$, denoted by $x_i$,
is
\begin{equation}
x_i=\frac{X_i}{\sum_{\ell\in A\cup B}X_\ell},
\end{equation}
whence we have
\begin{equation}
\dot{x}_i=
\frac{\dot{X}_i}{\sum_{\ell\in A\cup B}X_\ell}
-x_i\frac{\sum_{\ell\in A\cup B}\dot{X}_\ell}{\sum_{\ell\in A\cup B}X_\ell}=
\frac{\dot{X}_i}{\sum_{\ell\in A\cup B}X_\ell}-x_i\phi,
\label{eq:x*}
\end{equation}
where
\begin{eqnarray}
\lefteqn{\phi=
\sum_{jk\in A\cup B\atop j\neq k}f_jf_k\min\{x_j,x_k\}+}\hspace{1in}\nonumber\\
&&\sum_{jk\in A\cup B\atop j=k}f_j^2\frac{x_j}{2}+
\lambda\sum_{j\in A\cup B}f_jx_j.
\end{eqnarray}

From Eq.~(\ref{eq:x*}) we can easily derive the counterparts of
Eqs.~(\ref{eq:XA}) and~(\ref{eq:XB}) for relative abundances:
\begin{equation}
\dot{x}_i=\sum_{jk\in I_i\atop j\neq k}f_jf_kq_{jk,i}\min\{x_j,x_k\}+
\sum_{jk\in I_i\atop j=k}f_j^2q_{jj,i}\frac{x_j}{2}-
x_i\phi
\label{eq:xA}
\end{equation}
for $i\in A$ and
\begin{equation}
\dot{x}_i=\lambda\sum_{j\in I_i}f_jq_{j,i}x_j-x_i\phi
\label{eq:xB}
\end{equation}
for $i\in B$.

\subsection{Summary}

A summary of the topological and dynamical properties of hypergraph $H$ is given
in Table~\ref{table1}. In this table, emphasis is placed on the structure
of each type of hyperedge, as well as on how the existence of each hyperedge and
the associated dynamics depend on the model's parameters.

\begin{table*}[t]
\caption{Summary of the hyperedges of $H$ in relation to the model's parameters
and dynamics.}
\label{table1}
{\advance\textwidth-6\tabcolsep
\begin{tabular}{>{\raggedright}p{0.22\textwidth}>{\raggedright}p{0.22\textwidth}>{\raggedright}p{0.56\textwidth}}
\hline
Hyperedge's source multiset $S$			&Hyperedge's destination multiset $D$		&Properties\tabularnewline
\hline
Genotype set $\{j\}$\newline for $j\in A\cup B$		&Genotype set $\{i\}$\newline for $i\in B$		&\raggedright The existence of a hyperedge from $S$ to $D$ depends on probability $r$.\newline The dynamics, which represents the generation of genotype $i$ from mitosis and mitochondrial mutation at genotype $j$, depends on $r$, on the fitness $f_j$ of genotype $j$, and on the rate $\lambda$.\tabularnewline
\hline
Genotype multiset $\{j,k\}$\newline for $j,k\in A\cup B$	&Genotype set $\{i\}$\newline for $i\in A$	&The existence of a hyperedge from $S$ to $D$ depends on probabilities $p$ and $r$.\newline The dynamics, which represents the generation of genotype $i$ from meiosis with recombination at genotypes $j,k$, as well as on mitochondrial mutation at either $j$ or $k$, depends on $p$ and $r$, and on the fitnesses $f_j$ and $f_k$ of genotypes $j$ and $k$, respectively.\tabularnewline
\hline
\end{tabular}
}

\end{table*}

\subsection{A special case}
\label{special}

By Eq.~(\ref{eq:pmei}), for $p=r=1$ we have $p_{jk,i}=1$ for all $i\in A$
and all $j,k\in A\cup B$, meaning that every genotype $i\in A$ has the same
input set $I_i$. By Eq.~(\ref{eq:pmit}), for $r=1$ we similarly have
$p_{j,i}=1$ for all $i\in B$ and all $j\in A\cup B$, provided the ncDNA of
genotype $j$ is identical to that of $i$. In general, then, genotypes in $B$ can
have distinct input sets. In specifying the special case of this section, we aim
to study the setup in which not only every possible connection is present but
also the value of $X_i$ is the same for all $i\in A$ and likewise for all
$i\in B$. This can be imposed for $t=0$, but having it hold subsequently
requires moreover that every genotype keep receiving the same input as all
others in the same set ($A$ or $B$). This is already true of genotypes in $A$
(by virtue of their shared input set), but making it true of genotypes in $B$ as
well requires that we circumvent the fact that, inside a group of same-ncDNA
genotypes, fitness can be distributed differently than inside another group. We
can circumvent this by constraining the nature of the genotypes that constitute
sets $A$ and $B$ so that fitness distribution is the same for every occurring
ncDNA.

Our criterion for including genotypes in $A$ and $B$ is the simplest possible:
genotype $i$ is to be included if and only if both $i_1$ and $i_2$ are sequences
of $0$'s, so now all genotypes in $B$ have identical input sets as well. The
number of genotypes in $A$ or $B$ is therefore no longer the $G$ of
Eq.~(\ref{eq:G}), but instead $G_0=2^L$ (the number of possibilities for
mtDNA). Likewise, the number of genotypes having fitness $2^{-k}$, previously
given by the $n_k$ of Eq.~(\ref{eq:nk}), now amounts to
$n_{k,0}={L\choose k}$ (since $k$ is now the number of loci at which mtDNA is
$1$). Consequently, the sum total of fitnesses in set $A$ or $B$, previously
given by $F$ as in Eq.~(\ref{eq:F}), is now denoted by $F_0$ and given by
\begin{equation}
F_0=\sum_{k=0}^Ln_{k,0}2^{-k}=\left(\frac{3}{2}\right)^L.
\end{equation}

By Eqs.~(\ref{eq:qmit}) and~(\ref{eq:qmei}), $q_{k,i}=q_{jk,i}=1/G_0$.
Letting $a$ stand for any $i\in A$ and $b$ for any
$i\in B$, we obtain
\begin{equation}
\dot{X}_a=
\frac{F_0^2}{2G_0}X_a+
\frac{F_0^2}{G_0}\min\{X_a,X_b\}+
\frac{F_0^2}{2G_0}X_b
\end{equation}
and
\begin{equation}
\dot{X}_b=\frac{\lambda F_0}{G_0}X_a+\frac{\lambda F_0}{G_0}X_b.
\end{equation}
Rescaling time by the factor $F_0^2/2G_0$ and letting $\mu=\lambda/F_0$ allow us
to rewrite these equations as
\begin{equation}
\dot{X}_a=X_a+2\min\{X_a,X_b\}+X_b
\end{equation}
and
\begin{equation}
\dot{X}_b=2\mu X_a+2\mu X_b.
\end{equation}
Except for the trivial case of $X_i=0$ all over $A\cup B$ initially, these two
differential equations entail an unbounded exponential growth of both $X_a$ and
$X_b$.

There are two regimes to be considered. The first one, valid while $X_a\le X_b$,
leads to System~1,
\begin{eqnarray}
\dot{X}_a&=&3X_a+X_b,\\
\dot{X}_b&=&2\mu X_a+2\mu X_b,
\end{eqnarray}
whose eigenvalues are
$u_\pm=\mu+\frac{3}{2}\pm\sqrt{\left(\mu+\frac{3}{2}\right)^2-4\mu}$. Assuming
$X_a(0)=0$ yields
\begin{equation}
\frac{X_a}{X_b}=\frac{1-e^{(u_--u_+)t}}{u_+-3-(u_--3)e^{(u_--u_+)t}},
\end{equation}
and therefore,
\begin{equation}
\lim_{t\to\infty}\frac{X_a}{X_b}=\frac{1}{u_+-3}.
\label{ss1}
\end{equation}
This steady state can be reached only if it happens while $X_a\le X_b$, hence
for $\mu\ge 1$ ($\lambda\ge F_0$), with $\mu=1$ implying $X_a=X_b$. For $\mu<1$
the value of $X_a$ catches up with that of $X_b$ before the steady state can be
reached. In this case, the second regime takes over.

This second regime is valid for $X_a\ge X_b$ and leads to System~2,
\begin{eqnarray}
\dot{X}_a&=&X_a+3X_b,\\
\dot{X}_b&=&2\mu X_a+2\mu X_b,
\end{eqnarray}
of eigenvalues
$u_\pm=\mu+\frac{1}{2}\pm\sqrt{\left(\mu+\frac{1}{2}\right)^2+4\mu}$. For
$X_a(0)=X_b(0)$,
we obtain
\begin{equation}
\frac{X_a}{X_b}=
\frac
{3[(u_--4)e^{u_+t}+(4-u_+)e^{u_-t}]}
{(u_+-1)(u_--4)e^{u_+t}+(4-u_+)e^{u_-t}}
\end{equation}
and
\begin{equation}
\lim_{t\to\infty}\frac{X_a}{X_b}=\frac{3}{u_+-1}.
\label{ss2}
\end{equation}
Similarly to the first regime, this steady state can be reached only if it
happens while $X_a\ge X_b$, hence for $\mu\le 1$ ($\lambda\le F_0$), again with
$\mu=1$ implying $X_a=X_b$.

This brief analysis of Systems~1 and~2 reveals how $X_a>X_b$ can be reached in
the long run, having started at $X_a(0)=0$. The general picture is that the
genotypes are initially subject to the solution to System~1, which remains the
case for as long as $X_a\le X_b$. This can either endure indefinitely, with the
genotypes eventually reaching the steady state of Eq.~(\ref{ss1}), or end
when $X_a=X_b$ occurs before that steady state is reached. The outcome depends
on the value of $\lambda$, with $\lambda\ge F_0$ implying the former,
$\lambda<F_0$ implying the latter. It follows that, in order for the genotypes
to go beyond $X_a=X_b$ and eventually reach a steady state in which $X_a>X_b$,
we must have $\lambda<F_0$. In this case the genotypes become subject to the
solution to System~2 as soon as $X_a=X_b$ happens, and from then on converge to
the steady state of Eq.~(\ref{ss2}), necessarily with $X_a>X_b$. Note that
the key to this development from $X_a(0)=0$ is assigning a sufficiently small
value to $\lambda$. As we show in Sec.~\ref{sec:results}, this continues to
hold when we leave the special case and return to the general model.

\section{Results}
\label{sec:results}

Henceforth we use $x_A$ to denote the total relative abundance of genotypes
generated by meiosis, and likewise $x_B$ for those generated by mitosis. That
is, $x_A=\sum_{i\in A}x_i$ and $x_B=1-x_A$. All our results come from
time-stepping Eqs.~(\ref{eq:xA}) and~(\ref{eq:xB}) for a fixed instance of
hypergraph $H$, assuming $x_i(0)$ to be selected uniformly at random for
$i\in B$ and $x_i(0)=0$ for $i\in A$ (hence $x_A(0)=0$ and $x_B(0)=1$), then
averaging or binning over hypergraphs and initial conditions. Assuming total
dominance of mitosis at $t=0$ allows us to use our model to probe the
hypothetical evolutionary past in which cellular division by mitosis was the
rule but through mutation and natural selection gave rise to the appearance of
meiosis.

Unlike our previous work based on similar dynamical equations on random
networks \cite{bds12,bds15,bds16}, in which a few thousand nodes could be
handled within reasonable bounds on computational resources, the situation is
now significantly more severe. The reason behind this is twofold, since not only
does it involve a much faster growth of the number of nodes with sequence length
$L$ ($2G=2^{3L}+2^{2L}$ now versus $2^L$ or $2^{L+1}$ before), but also it
requires hyperedges (rather than edges) to be handled. As a consequence, in this
paper we use $L=3,4$, whereas in those previous publications we reached $L=10$
or even a little higher.

These limitations notwithstanding, we have been able to elicit richly varied
behavior from suitable valuations of the three parameters, $p$, $r$, and
$\lambda$. Our results are presented in Figs.~\ref{fig3}--\ref{fig7}. In
Fig.~\ref{fig3} we explore the parameter landscape as each possibility leads
to a form of coexistence between mitosis- and meiosis-generated genotypes. We do
this for both $L=3$ and $L=4$.

Figures~\ref{fig4}--\ref{fig6}, in turn, focus on how the relative abundances of
genotypes become distributed in the steady state, given a scenario in which
mitosis dominates ($x_A\approx 0.04$, in Fig.~\ref{fig4}), another in which
mitosis and meiosis coexist at roughly the same proportion ($x_A\approx 0.5$, in
Fig.~\ref{fig5}), and another in which meiosis dominates ($x_A\approx 0.96$,
in Fig.~\ref{fig6}). All three figures share the same value of $p$ and $r$,
with the value of $\lambda$ being set so that the desired steady-state value of
$x_A$ is reached. These figures are relative to $L=4$ only, but contain a lot of
further detail in that they include separate histograms for each of the $L+1$
possible values of a genotype's fitness ($2^{-L}$ through $2^0$).

Figure~\ref{fig7} is given in the same spirit of the previous three, now
containing plots for both $L=3$ and $L=4$. All its panels refer to the
eventual preponderance of meiosis over mitosis ($x_A\approx 0.96$) for fixed
$p$, with $\lambda$ adjusted to support the desired steady-state value of $x_A$
while $r$ is varied.

\section{Discussion}
\label{sec:disc}

\begin{figure}[t]
\includegraphics[scale=0.32]{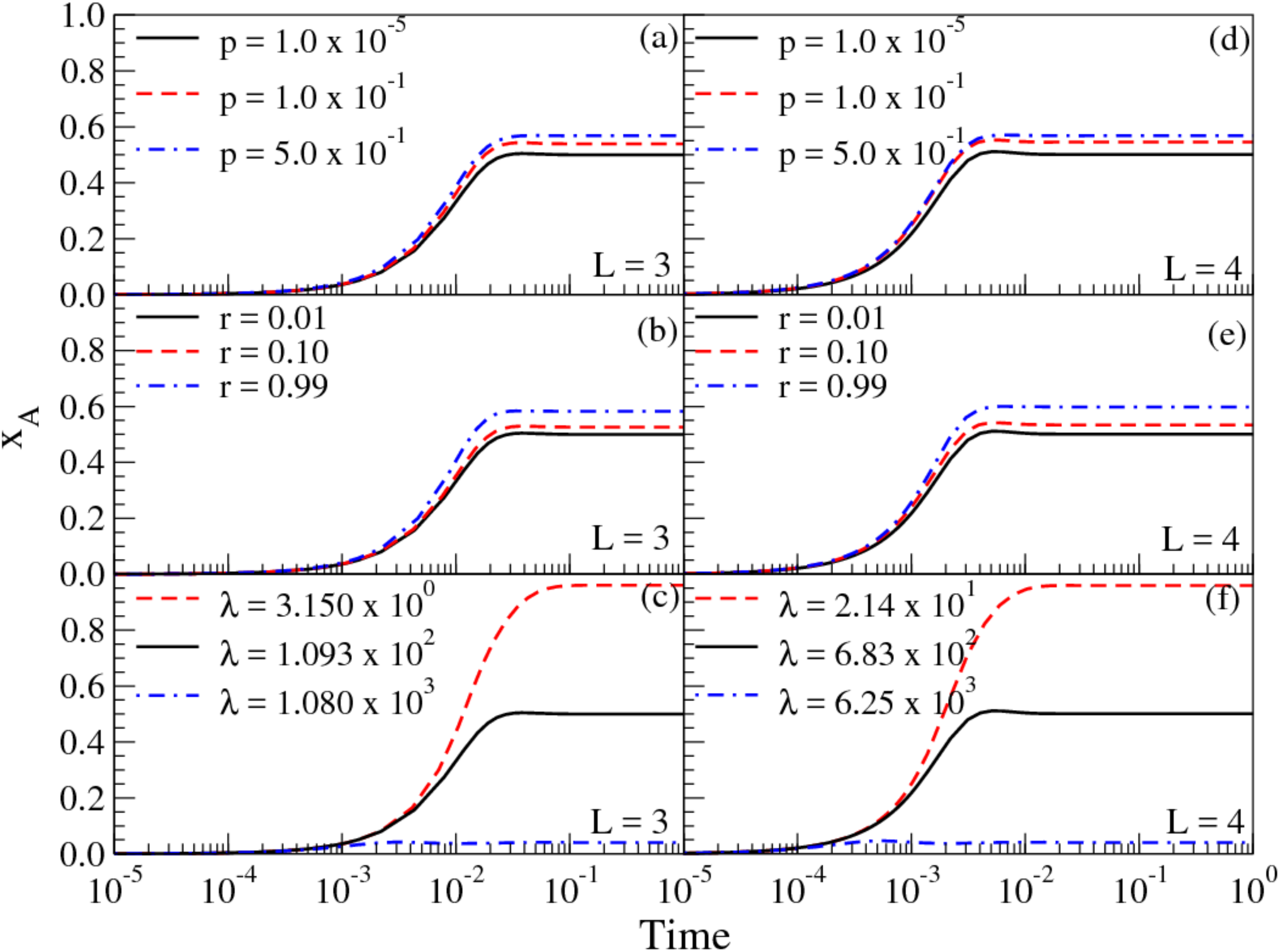}
\caption{Evolution of the relative abundance of genotypes generated by meiosis
($x_A$) for $L=3$ (a)--(c) and $L=4$ (d)--(f). Parameter values omitted from
each panel default to $p=10^{-5}$, $r=0.01$, and $\lambda=109.3$ (for $L=3$) or
$\lambda=683$ (for $L=4$).}
\label{fig3}
\end{figure}

Probabilities $p$ and $r$ influence network structure by affecting the
probabilities $p_{j,i}$ and $p_{jk,i}$ that each hyperedge exists: decreasing
$p$ leads to sparser meiotic hyperedges, decreasing $r$ leads to both sparser
mitotic hyperedges and sparser meiotic hyperedges. They also influence network
dynamics on a fixed hypergraph, both through the sparseness of the hyperedges
incoming to any given genotype and by relativizing these hyperedges' importance
(since now the normalized versions of $p_{j,i}$ and $p_{jk,i}$, respectively
$q_{j,i}$ and $q_{jk,i}$, are the ones in charge). Even though we might expect
the eventual steady state to depend on how $p$ and $r$ relate to each other, the
plots in Fig.~\ref{fig3} indicate that, in fact, it depends much more strongly
on the value of $\lambda$.

As explained in Sec.~\ref{netdyn}, $\lambda$ is intended to complement the
effect of squared fitnesses in lowering the rate of genotype production by
meiosis when compared to that of mitosis (to which fitnesses contribute
linearly). We see in Fig.~\ref{fig3} that, for $p=10^{-5}$ and $r=0.01$, using
$\lambda=3.15$ (for $L=3$) or $\lambda=21.4$ (for $L=4$) ensures the eventual
prevalence of meiosis over mitosis, while increasing these values slowly
reverses the trend and eventually leads to the prevalence of mitosis in the
steady state (panels (c) and (f)). Moreover, for the values of $\lambda$ leading
mitosis and meiosis to coexist approximately in the same proportion (these are
the default values for $\lambda$ in the figure), varying $r$ while $p$ remains
fixed at $10^{-5}$ (panels (b) and (e)) or varying $p$ while $r$ remains fixed
at $0.01$ (panels (a) and (d)) is practically innocuous. This remains true if
the other values of $\lambda$ in panels (c) and (f) are used instead (data not
shown), suggesting that, even though $p\ll r$ is known to hold currently, it
need not have held all along the evolutionary process. This obviates the need
for Assumption~1 (see section \ref{sec:intro}), at least as far as predicting
the eventual situation of mitosis- versus meiosis-generated genotypes is
concerned. In fact, this is true of Assumption~2 as well, since pushing the
value of $r$ very near $1$ ($r=0.99$) is practically tantamount to eliminating
$r$ from both Eq.~(\ref{eq:pmit}) and~Eq.~(\ref{eq:pmei6}) and yet
produces no relevant effect. The fact that the two assumptions are unnecessary
in the face of our results by no means invalidates the spirit of the hypothesis
we set out to explore in the first place, namely, that by evolving in
combination with mtDNA, the inheritance of ncDNA through cellular division by
mitosis can give rise to meiosis as the prevalent mechanism. In this case, the
mitochondria are seen to act not as predicated by Assumptions~1 and~2, but by
strongly influencing a genotype's fitness as well as the  multitude of pathways
through which evolution can work.

\begin{figure}[t]
\includegraphics[scale=0.32]{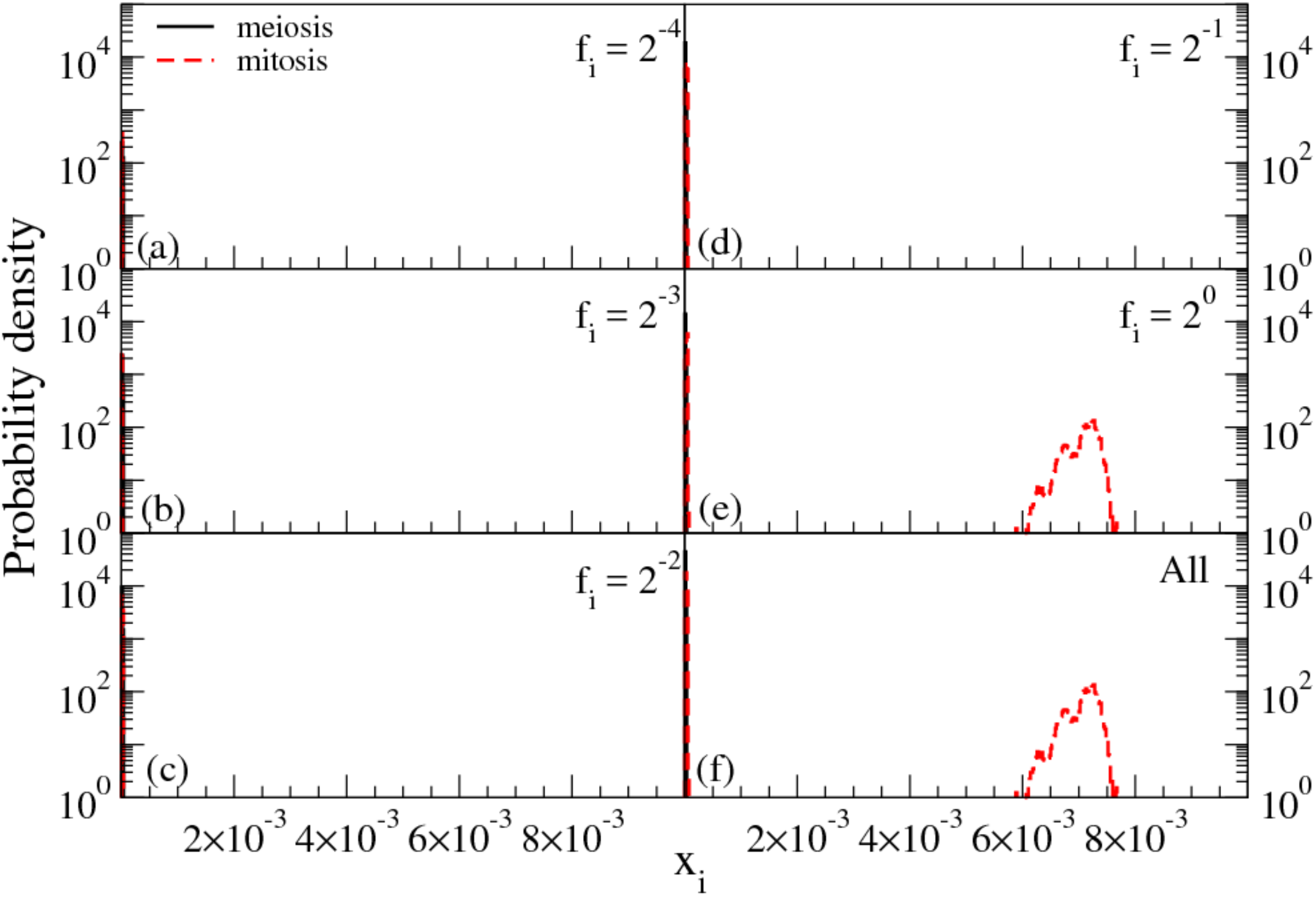}
\caption{Probability density of the stationary-state relative abundance of
genotype $i\in A\cup B$ for $L=4$ and $x_A\approx 0.04$ (i.e., mitosis
dominates), with $p=10^{-5}$, $r=0.99$, and $\lambda=2.84\times 10^3$. Each of
panels (a)--(e) refers to genotypes having the same fitness $f_i=2^{-k}$, as
indicated. Panel (f) refers to all genotypes. All panels contain density spikes
very near $x_i=0$ for both mitosis- and meiosis-generated genotypes.}
\label{fig4}
\end{figure}

\begin{figure}[t]
\includegraphics[scale=0.32]{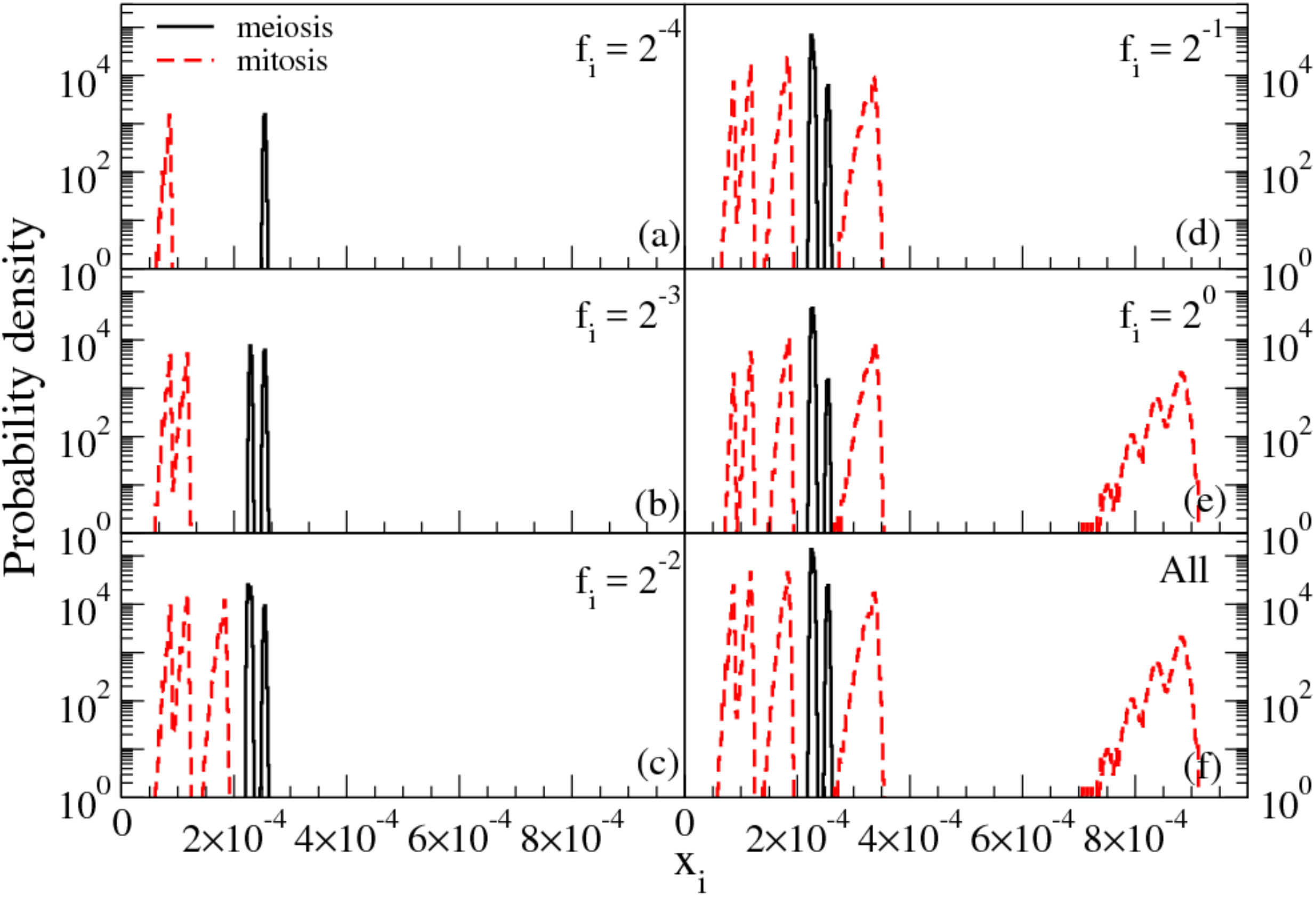}
\caption{Probability density of the stationary-state relative abundance of
genotype $i\in A\cup B$ for $L=4$ and $x_A\approx 0.5$ (i.e., mitosis and
meiosis coexist at about the same proportion), with $p=10^{-5}$, $r=0.99$, and
$\lambda=938$. Each of panels (a)--(e) refers to genotypes having the same
fitness $f_i=2^{-k}$, as indicated. Panel (f) refers to all genotypes.}
\label{fig5}
\end{figure}

\begin{figure}[t]
\includegraphics[scale=0.32]{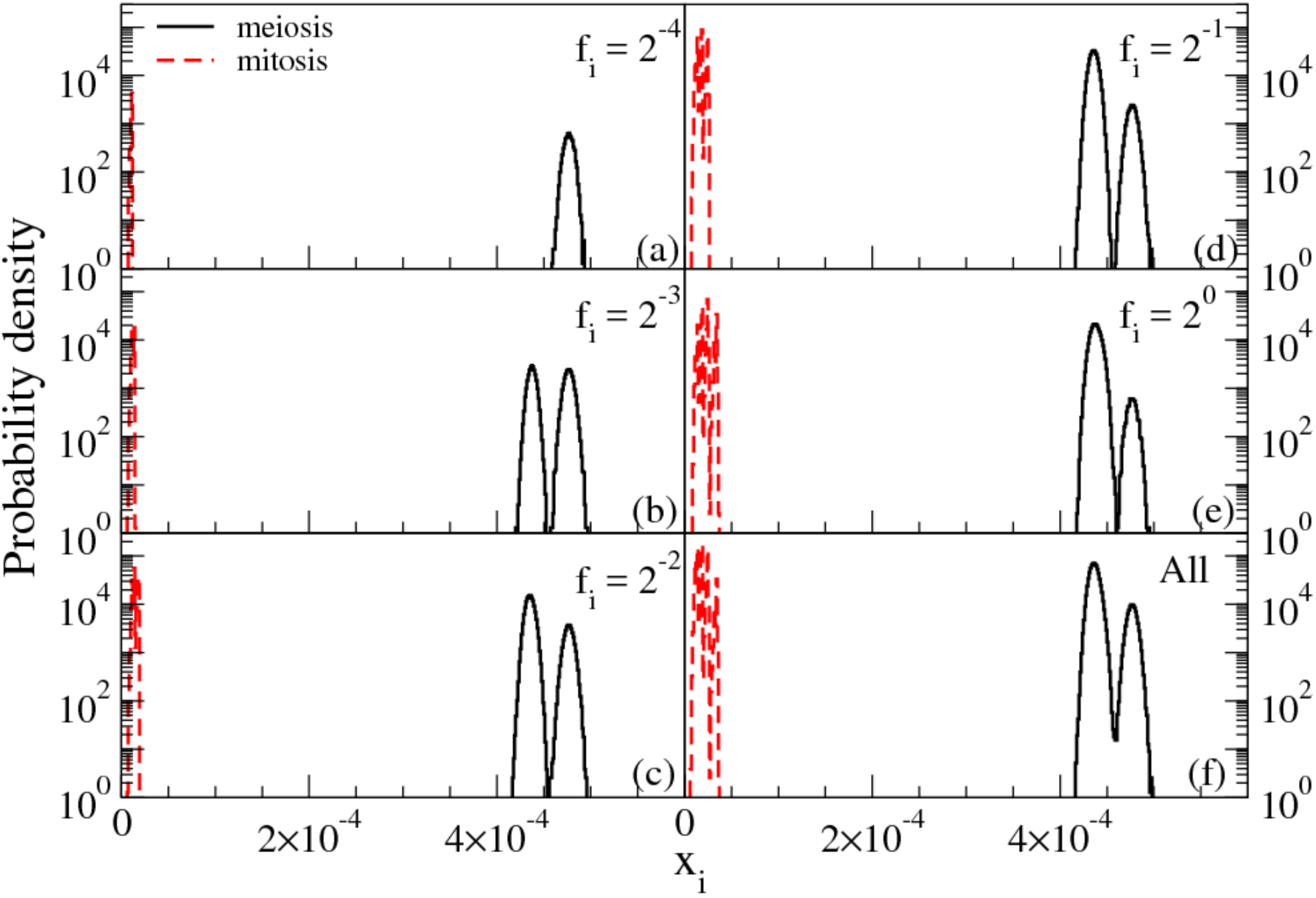}
\caption{Probability density of the stationary-state relative abundance of
genotype $i\in A\cup B$ for $L=4$ and $x_A\approx 0.96$ (i.e., meiosis
dominates), with $p=10^{-5}$, $r=0.99$, and $\lambda=26.5$. Each of panels
(a)--(e) refers to genotypes having the same fitness $f_i=2^{-k}$, as indicated.
Panel (f) refers to all genotypes. Panels (a)--(c) contain density spikes very
near $x_i=0$ for mitosis-generated genotypes.}
\label{fig6}
\end{figure}

Of all the $p/r$ ratios used in Fig.~\ref{fig3}, we use the lowest
($p=10^{-5}$ to $r=0.99$) in Figs.~\ref{fig4}--\ref{fig6}, which in sequence
refer to a progression in the eventual relative abundance of genotypes generated
by meiosis (from very low in Fig.~\ref{fig4} to very high in
Fig.~\ref{fig6}, with $\lambda$ adjusted accordingly all along). These figures
refer to $L=4$ and show the probability density of relative genotype abundances
for each possible fitness value. By Eq.~(\ref{eq:nk}), $16$ of the
mitosis-generated genotypes have the least possible fitness ($2^{-4}$), followed
by $128$, $480$, $896$, and $656$ genotypes as we move higher up. The same holds
for the meiosis-generated genotypes.

What we see in Fig.~\ref{fig4} (which corresponds to the meiosis-generated
genotypes reaching only a small relative abundance in the steady state,
$x_A\approx 0.04$) is that only the very fittest of the mitosis-generated
genotypes survive, that is, only about $30\%$ of them. This indicates that
co-evolution with genotypes generated by meiosis exerts considerable pressure
on those generated by mitosis even in conditions that are highly favorable to
the latter. As the value of $\lambda$ gets decreased to support $x_A\approx 0.5$
(Fig.~\ref{fig5}), we start to see nonzero relative abundances at all fitness
levels for both mitosis- and meiosis-generated genotypes. Nevertheless, some of
the fittest mitosis-generated genotypes continue to dominate, occurring with the
highest relative abundances in isolation, though at one full order of magnitude
lower than previously. What supports $x_A\approx 0.5$ in the steady state even
in these conditions is the fact that meiosis-generated genotypes are already
winning at the three lowest fitness levels (roughly $29\%$ of such genotypes),
as well as the clash between mitosis- and meiosis-generated genotypes at the
following fitness level (about $41\%$ of the genotypes in each group).
Decreasing $\lambda$ considerably further leads to the steady-state scenario
shown in Fig.~\ref{fig6}, which corresponds to the rise of the
meiotic-generated genotypes to dominance ($x_A\approx 0.96$). Readily, genotypes
at all fitness levels contribute, which contrasts sharply with
Fig.~\ref{fig4}.

\begin{figure}[t]
\includegraphics[scale=0.32]{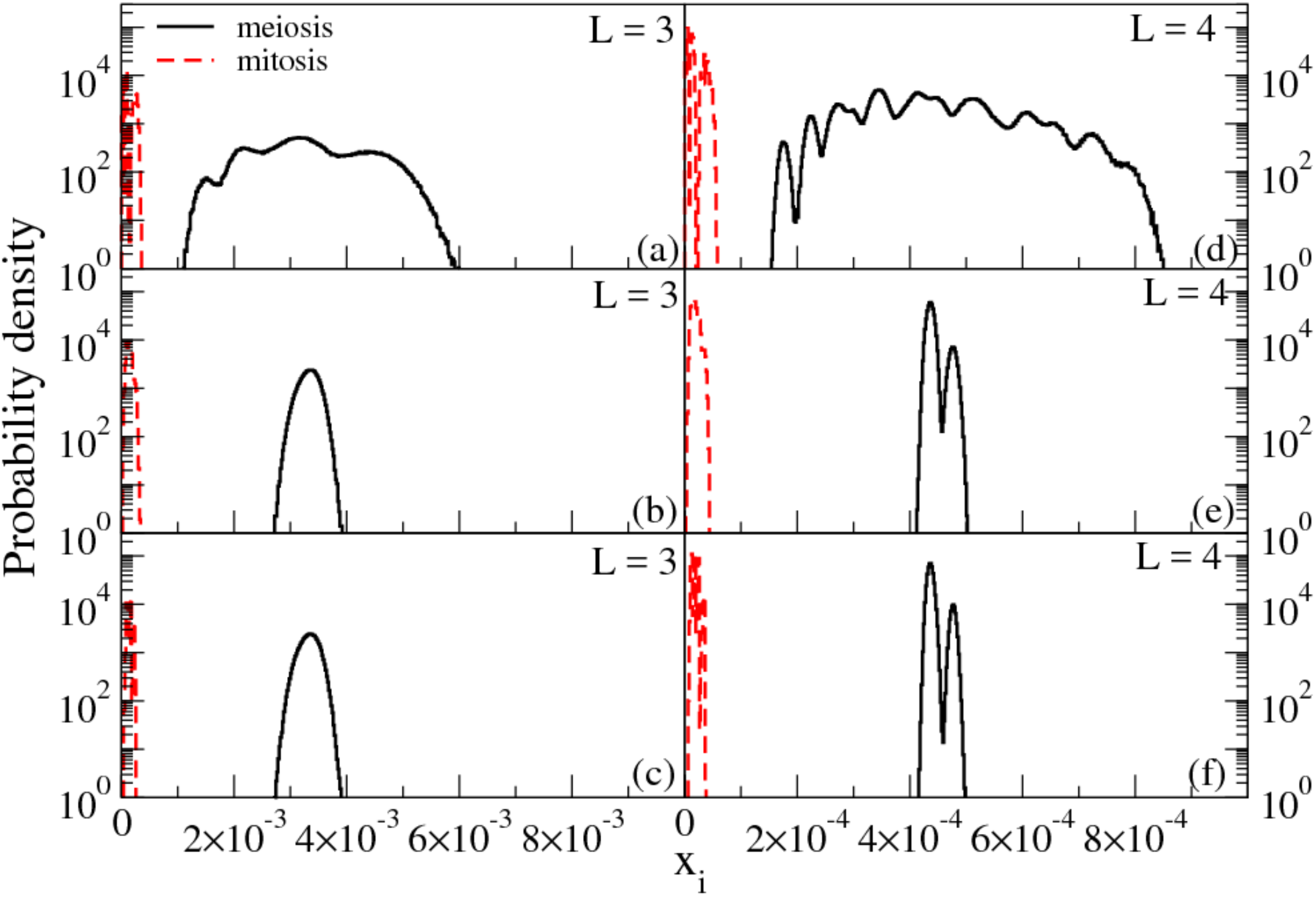}
\caption{Probability density of the stationary-state relative abundance of
genotype $i\in A\cup B$ for $L=3$ (a)--(c) and $L=4$ (d)--(f), in both cases for
$x_A\approx 0.96$. Probability $p$ is fixed at $p=10^{-5}$; probability $r$
varies from $r=0.01$ (a), (d), to $r=0.9$ (b), (e), to $r=0.99$ (c), (f); the
value of $\lambda$ for each panel is as follows: $\lambda=3.15$ (a),
$\lambda=4.1$ (b), $\lambda=4.05$ (c), $\lambda=21.4$ (d), $\lambda=27.7$ (e),
$\lambda=27.7$ (f). Panel (f) is identical to Fig.~\ref{fig6}(f). Panels
(a)--(c) contain density spikes very near $x_i=0$ for mitosis-generated
genotypes.}
\label{fig7}
\end{figure}

A further view of the probability densities associated with the relative
abundance of genotypes is given in Fig.~\ref{fig7}, aiming to complement
Figs.~\ref{fig4}--\ref{fig6} by including some $L=3$ cases and by varying the
value of $r$. The plots in Fig.~\ref{fig7} are all given for $x_A\approx 0.96$
in the steady state, a setting similar to that of Fig.~\ref{fig6}, now letting
$r$ vary from $0.01$ (a), (d), to $0.9$ (b), (e), to $0.99$ (c), (f). As before,
the value of $\lambda$ has been calibrated to yield the desired $x_A$. As
mentioned earlier, increasing the value of $r$ leads to denser networks in
relation to both the hyperedges leading to mitosis-generated genotypes and to
those leading to meiosis-generated genotypes. As shown in Fig.~\ref{fig7}, the
effect of this is to progressively narrow the interval where nonzero densities
are found, particularly so for the winning variety, that of meiosis-generated
genotypes. This seems to be indicating that denser networks tend to homogenize
relative abundances across genotypes generated in the same manner, at least to
a certain extent.

To finalize, we return to the special case of Sec.~\ref{special}, where we
found $F_0$ to be a strict upper bound on the value of $\lambda$ in order for
meiosis-generated genotypes to prevail over mitosis-generated ones. This
indication that lowering the value of $\lambda$ in the special case is crucial
to the prevalence of meiosis-generated genotypes is true of the general case as
well, as demonstrated in Fig.~\ref{fig3}. In fact, the analogue of
$\lambda<F_0$ in the general case is $\lambda<F$, which as we see in
Figs.~\ref{fig3}(c),~(f),~\ref{fig6}, and~\ref{fig7} (those in which the
prevalence of meiosis over mitosis occurs), holds as well: by
Eq.~(\ref{eq:F}), we have $F=185$ for $L=3$, $F=1\,241$ for $L=4$,
therefore substantially higher than the values of $\lambda$ used in those
figures. We find it remarkable that such constraint on the value of $\lambda$
should remain essentially valid all the way up from the drastically simplified
case of Sec.~\ref{special}.

\section{Conclusion}
\label{sec:concl}

Our network model of how the mitotic and meiotic modes of cellular division may
have co-evolved, eventually giving rise to meiosis as the prevalent mechanism
underlying reproduction in eukaryotes, includes an oversimplified representation
of genotypes and only three parameters. Of these, two (probabilities $p$ and
$r$) aim to account for the stochastic variability in DNA during cellular
division and the third ($\lambda$) aims to adjust the growth rate of genotype
populations generated by mitosis to that of genotype populations generated by
meiosis. Probability $p$ is related to the recombination that ncDNA undergoes
during meiosis, while probability $r$ is related to mutations that mtDNA
undergoes when transmitted from parent cell to offspring and is therefore
present in division by both mitosis and meiosis. 

In spite of its simplicity, we have found the model to yield surprisingly
complex behavior and to contemplate a variety of scenarios regarding the
steady-state coexistence of the two modes of cellular division. In particular,
we have identified values of $\lambda$ for which the genotypes generated by
meiosis rise to dominance from initial conditions in which those generated by
mitosis dominate absolutely, thus providing theoretical support to the most
recent hypothesis as to why eukaryote reproduction has come to be dominated by
meiosis as the most common mode of cellular division. Recall from
Sec.~\ref{sec:intro} that at the core of this hypothesis is the random
diversification of ncDNA afforded by meiosis as a means to catch up with that of
mtDNA. Our three parameters have been meant precisely to capture this mismatch.
Together with the various model details, they provide a mathematically sound
view into what may have happened along evolutionary time.

However, our model's support to the said hypothesis is not without its caveats.
Even though the model's two main assumptions (Assumptions~1 and~2), given in
Sec.~\ref{sec:intro}, place heavy weight on $p\ll r$ and on the inheritance
of mtDNA solely from one of the two parents when cellular division happens by
meiosis, our results indicate that neither of the two assumptions is essential.
Instead, they suggest that, in association with mitochondria, cellular division
by meiosis is naturally poised to navigate a fitness landscape heavily
influenced by how the two forms of DNA interact much more efficiently than
cellular division by mitosis.

Our conclusions still lack in robustness, though. Whether they will stand
further stressing depends on our future ability to extend our computational
results to substantially lengthier genotypes. Exactly how to achieve this will
be the subject of further research.

\begin{acknowledgments}
We acknowledge partial support from Conselho Nacional de Desenvolvimento
Cient\'\i fico e Tecnol\'ogico (CNPq), Coordena\c c\~ao de Aperfei\c coamento de
Pessoal de N\'\i vel Superior (CAPES), and a BBP grant from Funda\c c\~ao Carlos
Chagas Filho de Amparo \`a Pesquisa do Estado do Rio de Janeiro (FAPERJ).
We also acknowledge support from Agencia Nacional de Investigaci\'on e
Innovaci\'on (ANII) and Programa de Desarrollo de las Ciencias B\'asicas
(PEDEClBA). We thank N\'ucleo Avan\c cado de Computa\c c\~ao de Alto Desempenho
(NACAD), Instituto Alberto Luiz Coimbra de P\'os-Gradua\c c\~ao e Pesquisa em
Engenharia (COPPE), Universidade Federal do Rio de Janeiro (UFRJ), for the use
of supercomputer Lobo Carneiro, where part of the calculations was carried out.
\end{acknowledgments}

\bibliography{mxxosis}
\bibliographystyle{apsrev}

\end{document}